\documentclass[paper=a4]{article}
\pdfoutput=1
\usepackage[english]{babel}

\usepackage{a4}
\usepackage{ucs}
\usepackage[utf8x]{inputenc}
\usepackage[T1]{fontenc}
\usepackage{graphicx}
\usepackage{amssymb}
\usepackage{longtable}

\newcounter{definition}

\begin{document}

\title{Two limitations of our knowledge of quality}
\author{Johannes Reich, johannes.reich@sap.com}

\maketitle

\begin{abstract} This article develops a quality notion that is complementary to the system notion. As a major consequence, it becomes clear why quality can be measured only to a certain extend based on the issues of validity and incompleteness. First, there is an inherent conflict between the applicability and validity of quality measures and second, quality considerations almost always refer to high-dimensional spaces with only sparse knowledge also known as "curse of dimensionality". The resulting gap of knowledge has to be filled by experienced based heuristics. To deal with the curse of dimensionality, the heuristics of categorizing qualities into strategic and necessary is proposed. Strategic qualities provide contrast, while necessary qualities rather diminish contrast. In an economic context the presence of strategic qualities motivate a buy-decision and the absence of necessary qualities motivate a don't-buy-decision.   
\end{abstract}

%
\section{Introduction}
%
 
{\it ''There is nothing more practical than a good theory''}, a saying that is attributed to James Clerk Maxwell\footnote{Some authors attribute it to Ludwig Boltzmann or Kurt Lewin.} Judging the contradictory impressions the current literature on product quality still provides, one might conclude that a bit more theory would do well. Actually, already Rui S. Sousa and Christopher A. Voss, in their review about quality management \cite{SousaVoss2002_Quality}, saw a need for a stronger theory of quality (management).

First, there is an obvious discrepancy between the mantra-like proposition that measuring quality is essential for any quality control and the deficit of the myriad of existing quality models to only rudimentarily define meaningful measurable factors. Using a survey, Stefan Wagner et al. \cite{WagnerEtAl2012_Software} found that 70\% or their respondent reported to use a company-specific quality model, 28\% used ISO/IEC 9126 and only 4\% ISO/IEC 25000. They state that {\it ''operationalizing quality models was one of the major problems encountered in practice. In the existing quality models, it was perceived that concrete measures that are needed for the operationalization are missing.''}.

This gap is emphasized by the seemingly contradictory position that quality has to part of a company's management culture and values - categories which should not be worth mentioning, if quality were as measurable as is often proposed.

Also, it is often taken for granted that quality relates only to system ''inherent'' attributes expressing a lack of understanding how different classes of properties come about. 

As a result, many approaches dealing with product quality seem to be an amalgam of possible sensible heuristics and pre-scientific schools-thinking, resulting in a surprisingly sparse state of facts with respect to their effectiveness, despite their ubiquitous dissemination. A good example is TQM. In their 2001 literature survey, Carlos J.F. Cândido and Sérgio P. Santos \cite{CandidoSantos2011_TQM} state that it was very difficult to present an estimate for the likelihood of TQM success as the reported rates of failure in their literature survey ranged from as low as 7\% to as high as 80\%. Approximately 10 years later, Ali M. Mosadeghrad \cite{Mosadeghrad2012_Towards} states that, with a focus on the special characteristics of the health sector states, {\it ''the literature provides little evidence of the effectiveness of TQM (in healthcare)''}.. 

The contribution of this article is to resolve these contradictions by providing a conceptual framework, a theory, as I chose to say, which allows for a better understanding on the role of quality in successful product development with some emphasis on software engineering. I present a quality notion that is based on the notion of a system and its environment.  As described in \cite{Reich2016_systems} a single relation between states\footnote{understood as functions from the time domain into the respective value domain} - the functionality - can be viewed as constituting a (technical) system. And complementary to that relation, all other relations of this system to its environment can be subsumed under the quality notion. So, functionality and quality play a complementary role. Roughly speaking, the function tells us ''what'' a system does and the qualities tell us ''how'' a system does so. Several classes of system properties and therefore qualities are proposed: compositional vs. emergent, explicit vs. implicit and systemal vs. interactional. The presented insights should be important for all business and technical people who devote themselves to the development of products with sufficient quality from a strategic perspective. 

With respect to our ability to objectify quality, two insurmountable obstacles are pointed out, namely incompleteness due to high dimensionality and the inherent conflict between validity and applicability, rendering practically complete knowledge about quality impossible. These aspects of quality enable a better understanding of its dichotomous role: on the one hand, what we know about quality can be used for strategic decisions on system development and on the other hand what we don't know about quality has to be bridged by experienced based heuristics, that is different techniques dealing with uncertainty. 

The structure of the article is the following: first, the quality notion is discussed in depth. Then the two principal reasons restricting the measurability of quality are introduced. Finally, the main results are summarized.

%
%
\section{The definition of quality}
%
In modern times, there have been numerous efforts to define quality in an engineering sense - with astonishingly diverse results. Quality is defined as a ''totality'', as a ''degree'', or as a ''capability'', to name just a few: 

\begin{itemize}
\item{\bf DIN EN ISO 8402:1995-08} The old norm of quality management defined quality as {\it ''The totality of characteristics of an entity that bear on its ability to satisfy stated and implied needs''}.
\item{\bf DIN EN ISO 9000:2005} The ISO 9000 family of standard on quality management systems defines quality as {\it ''The degree to which a set of inherent characteristics fulfills requirements''}.
\item{\bf ISO/IEC 25000:2005} The ISO/IEC 25000 family of standards for Software Product Quality Requirements and Evaluation (SQUaRE) defines software quality as the {\it ''Capability of software product to satisfy stated and implied needs when used under specified conditions''}. ISO/IEC 25020:2007 further distinguishes between external and internal software quality. 
\end{itemize}

Essentially most of the modern definitions entail the following three elements:
\begin{itemize}
\item Quality describes aspects of the relation between a system, parameterized by attributes, and  some (usage) context.
\item It is this context which poses the requirements for the manifestation of these attributes, that is, the assessment whether something is useful/good or not/bad.
\item And last, but certainly not least, we have the aspiration that any discussion of quality can be understood by everyone, or in other words, it is supposed to be objective, which means, among others, measurable. 
\end{itemize}

To understand quality, we therefore have to know what a system is, what usage means, what a context of usage is and what a measurement is. In the following, I will further elaborate on these topics.

\subsection{A system perspective}
Actually, as is stressed by many authors (e.g. \cite{Bertalanffy1968_GST,Sifakis2011_Vision}) the system notion bears the promise to provide a unifying view on such diverse disciplines as civil, mechanical, electrical, or software engineering - they all deal with the design of systems. 
In the world of electrical engineering, IEC 60050 \cite{IEC_60050} defines a system (351-42-08) as a ''set of interrelated elements considered in a defined context as a whole and separated from their environment''. It is noted that a system is generally defined with the view of achieving a given objective, for example by performing a definite function.

\begin{figure}[htbp]
  \begin{center}
    \includegraphics[width=6cm]{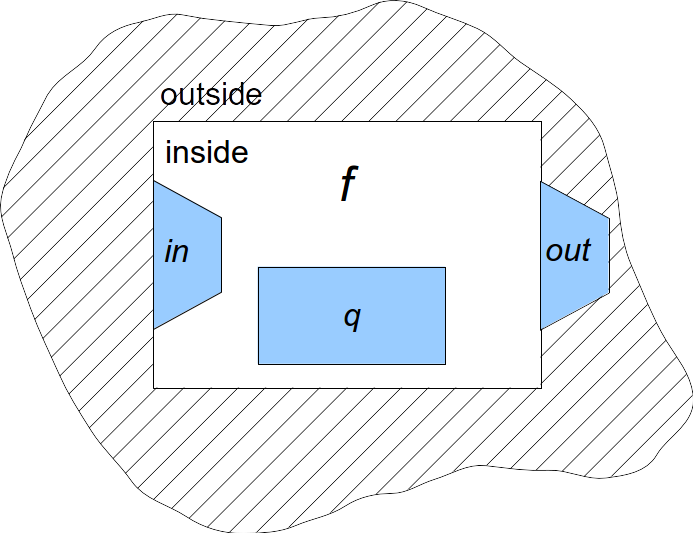}
\raisebox{1.5cm}{
\begin{tabular}[b]{p{2.9cm}|p{0.4cm}|p{0.4cm}|p{0.4cm}}\\
 & q & in & out\\
\hline
\hline
Input to system function      & $\surd$ & $\surd$ & - \\
\hline
Influenced by system function & $\surd$ & -       & $\surd$ \\
\hline
Belongs to several systems    & -       & $\surd$ & $\surd$  
\end{tabular}
}
 \end{center}
\caption[]{Diagram of a system. The state functions $(q, in, out):T \rightarrow Q\times I\times O$, mapping the time domain onto the sets of internal state values, input state values and output state values, become separated from the rest of the world by the system function $f:Q\times I\rightarrow Q\times O$, mapping the sets of the internal state values together with the input state values onto the sets of the internal state values together with the output state values. The asymmetric boxes of the input and output state symbolizes their asymmetric use in the system model. (Fig. drawn from \cite{Reich2016_systems})} \label{fig_system}
\end{figure}

So it seems to be consensus that a system separates the world into an inside and an outside, a system and its environment - suggesting the major question how this separation comes about. In \cite{Reich2016_systems} the idea is elaborated that the functional relation between the values of states at given points in time provide the system-constitutive property, separating these states of the system from the rest of the world, as is illustrated in Fig. \ref{fig_system}. Such a relation implies causality and time.

According to this view, system borders are drawn by fundamental mathematical structures and thereby may become very dynamic, as they are created with the creation of functionality and disappear together with it. As the language of mathematics is the common denominator of the diverse engineering disciplines, such a view may indeed live up to the promised unifying character of the system notion.

Depending on the class of system function, time, or state, different classes of systems can be identified. Important classes of functions are computable functions, finite functions or analytic functions. Important classes of times are discrete and continuous times and important classes of states are classical and quantum states.

But a system relates to the rest of the world – its environment – in many more important ways than just by its functionality. This is where quality comes into play to subsume all the other relations of a system to its environment, like weight, size, robustness, security, etc.

We can thereby say that both, quality and functionality play a truly complementary role. Functionality is about the “what” a system does and quality is about the “how” a system does so. Systems providing the same functionality can differ enormously in their way how they provide it. Or to put it more abstractly: while functionality is system-constitutive, quality is about the many and diverse non-functional relations between the system and its environment. Often the term ''Non-Functional Requirements'' is used to denote qualitative requirements. 

Depending on the functioning of ''its'' system, qualities often have a holistic character. There is, for example, not much sense in speaking about the performance or security of a broken system. 

\subsection{Compositionality}
Regardless of the differences in the details, any serious system model provides some concept on how systems compose to supersystems, that is about compositionality. As is discussed in \cite{Reich2016_systems}, basing the system notion on the system function allows for the following notion of compositionality of system properties\footnote{This approach follows an idea of Arend Rensink, proposed in his talk ''Compositionality huh?'' at the Dagstuhl Workshop ``Divide and Conquer: the Quest for Compositional Design and Analysis'' in December 2012.}

\begin{definition}\label{def_compositionality} Be $S$ the set of all systems and be $comp:S^n \rightarrow S$ a composition operator. A system property $\alpha$ is a partial function $S \rightarrow A$ which attributes values of some attribute set $A$ to systems ${\cal S}\in S$. 

A property $\alpha$ of a composed system ${\cal S} = comp({\cal S}_1, \dots, {\cal S}_n)$ is called  ''compositional'' if a function $comp_{\alpha}: A^n \rightarrow A$ exists that fulfills the following relation:

\begin{equation} 
\alpha\left(comp({\cal S}_1, \dots, {\cal S}_n)\right) = comp_\alpha\left(\alpha({\cal S}_1), \dots, \alpha({\cal S}_n)\right)
\end{equation}

Otherwise the system property is called ''emergent''.  
\end{definition}

In other words, compositional properties of the composed system result exclusively from the respective properties of the parts, emergent properties result also from other properties of the parts \cite{Reich2001}. A simple example of a compositional system property is the mass $m({\cal S})$ of a physical system ${\cal S}$. The total mass of the composed system is simply the sum of the masses of the single systems: $m\left(C({\cal S}_1, \dots, {\cal S}_n)\right) = m({\cal S}_1) + \dots + m({\cal S}_n)$. 

A simple example of an emergent property is the resonance frequency of an RLC-circuit. These properties are emergent and not compositional, simply because the building blocks don't show them. 

As qualities relate to system properties, it therefore makes sense to classify qualities also along the dimension of compositionality. Compositional qualities behave compositional in the sense of Def. \ref{def_compositionality}. An example could be the resource consumption of several components in the benign case. 

An example of an emergent quality certainly is security, where it is a well known that the security of a composed system generally does not solely depend on the security of its components.

\subsection{Explicit and implicit properties}

In analogy to the mathematical concept of explicit and implicit functions, I would like to distinguish between ''explicit'' and ''implicit'' system properties. An explicit property of a system can be represented by an explicit function, mapping the system time onto the property values and thereby can be validly represented by a state of a system. In the case of the RLC-circuit, the resonance frequency could be expressed as a function of the property values of the resistor, capacitor, and inductor.

An implicit property of a system is a property that generally cannot be explicitly represented by such a known function, but is nevertheless represented by an implicit function. Important examples are properties of while-systems \cite{Reich2016_systems} whose time steps are realized only when a given explicit computable function has a first zero in an iterative process. Another example is stability of nonlinear systems which could be characterized by a Lyapunow-function but no general algorithm exist to determine one \cite{GieslHafstein2015}.  

\subsection{Systemal versus relational properties}
Until now we have considered only properties that can be attributed to single systems. I call such a property ''systemal''. These kind of properties are probably denoted by the ISO 9000 family of standards as ''inherent'' attributes. But there are others - very important - properties which can only be defined with respect to some relation a system has to the rest of the world, that is to other systems, like an interaction. I propose to name such properties ''relational''. So, for a relational property, its not only the assessment of being good or bad that depends on the context, as is the case also for systemal properties, but the property itself has to be defined with respect to some context. 

An important example is security. Modern cryptography (see e.g. \cite{KatzLindell2015} for an overview) has made enormous progress since it was discovered how to define security with respect to some capabilities of an adversary, e.g. that she is able to use probabilistic polynomial-time algorithms and what information is available to her. Another good example for a relational quality is usability. Whether a given software is usable or not, not only depends on the system comprising of software and hardware, but also on the level of experience of the user as well as her expectations. 

%
\section{Objectify quality: quality measurement}
%

The role of the context (of usage) for the definition of quality requires the following three determination steps to make quality measurable:

\begin{enumerate}
\item the context (of usage) for the system under consideration,
\item the relevant attributes (of the system or its relations), and
\item the requirements with respect to these attributes.
\end{enumerate}

This enables to measure quality as the grade of fulfillment of the attributes compared to the requirements. So, the good news is: yes, with such a procedure we can measure quality. But, the bad news is: only to a certain extent. There are two insurmountable obstacles calling for an essential role of experienced based human intuition in our judgment on quality: incompleteness due to high dimensionality and the inherent conflict between validity and applicability

\subsection{Validity versus applicability of quality measurement}
The international standard's proposed approach to quality measurement is a hierarchical decomposition of both, the quality factors as well as the defined metrics or measures\footnote{In 1976 Tom Gilb coined the term ''software metrics'' in his book with the same title \cite{Gilb1976_Software}. With ISO/IEC 25000, this term was abandoned in favor of ''measure''}. The IEEE 1061-1998 Standard for a Software Quality Metrics Methodology \cite{IEEE_1061-1998} defines a ''direct metric'' as a metric that supposedly {\it ''does not depend upon a measure of any other attribute''}. It is proposed that each quality factor has an associated direct metric as a quantitative representation of this quality factor. Mean time to failure (MTTF) is given as an example for such a direct metric of the quality factor ''system reliability''. ISO/IEC 25020 \cite{ISO_IEC_25020} defines a ''base measure'' as a measure {\it ''defined in terms of an attribute and the method for quantifying it''} and states that a base measure {\it ''is functionally independent of other measures''}. A ''derived measure'' is a {\it ''measure that is defined as a function of two or more values of base measures''}\footnote{Additionally, a ''quality measure element'' is defined as the combination of the attribute [the measure], the measurement method and, optionally, the transformation by a mathematical function. But as the method for quantifying a measure is already part of the measure definition and it does not make much sense to distinguish between the measurement function and any subsequently applied extra function.}.

Cem Kaner and Walter P. Bond \cite{KanerBond2004_Software} provide an excellent discussion of the topic of construct validity with two in depth examples of ''Mean Time to Failure (MTTF)'' and ''bug count''. They show that even these simple measures are by no means ''direct'' or ''base'' in the proposed sense, but depend in their meaningfulness on many additional contextual assumptions. 

First, why mean? The arithmetic mean is a measure of location of a distribution. Another one is the median, others are geometric mean, the harmonic mean, or any one of the infinitely many generalized Hölder means. But often, a distribution of values is not sufficiently described by some location but also a measure of variation is needed, like the standard deviation in case of a Gaussian distribution. But other distributions require other measures of variations. Perhaps, the TTF does not even follow any prescribed distribution sufficiently exact at all. 

Cem Kaner and Walter P. Bond also note that there are many different ''times'': user time, calender time, processor time, etc. - as well as many different failure types: program crashes, data corruptions, error messages, help messages, any event that wastes some user time, etc. They even mention subjective factors like the individual's experience with the product or the user subpopulation's operational profile.

In my opinion, at the heart of this construct validity debate stands the fact that a measurement of a quality attribute value is by itself again a function, i.e. a unique mapping which has to include the context! And this context dependency results in a fundamental, that is unavoidable, conflict between validity and applicability of our quality measure. The narrower we make our context specification, the more valid the quality measurement function becomes for this special context but the less applicable it will be for other contexts. One could name it the ''semantic uncertainty relation''. As a result, quality notions which are supposed to be valid in a very broad context, like ''being successful'' are hardly measurable at all. 

\subsection{Incompleteness of quality measurement: the curse of dimensionality}
In their review on quality management \cite{SousaVoss2002_Quality}, Rui S. Sousa and Christopher A. Voss,  point our that the {\it ''importance of recognizing the multi-dimensional nature of quality cannot be overstated''}. The question is: exactly what kind of consequences does the high-dimensional nature of quality entails? 

Due to the infinite ways in which a system can relate to its environment we are hit by what Richard E. Bellman \cite{Bellman1961} called the ''curse of dimensionality'', the cumbersome, non-intuitive aspects of high-dimensionality. In the case of quality, the adversity between the infinite parameter space and the enforced limitations of our necessarily focused approach to quality leads to the non-intuitive effect that adding more and more dimensions does not necessarily help in differentiating better between the good and the ugly, but quite often does the opposite: the contrast diminishes. 

There have been numerous attempts to systematically classify these ''ilities'' and create taxonomies. Starting with the early models of Barry W. Boehm et al. \cite{BoehmEtAl1978_Characteristics} and J.A. McCall, P.K. Ritchards and G. F. Walters \cite{McCallRichardsWalters1977}, the currently most prominent one is the ISO/IEC 25000 standard family \cite{ISO_IEC_25000} with the quality model defined in ISO/IEC 25010 \cite{ISO_IEC_25010}.

Although the basic concepts are more or less similar, the different schemes are - not very surprisingly - inconsistent with one another (for an overview see \cite{AlQutaish2010_quality,ChungLeite2009}). Security, for example, was not put at top-level in ISO/IEC 9126 \cite{ISO_IEC_9126} which was changed in ISO/IEC 25010. The dependability classification by Algirdas Avizienis, Jean-Claude Laprie, Brian Randell, and Carl Landwehr \cite{AvizienisLaprieRandellLandwehr2004_Basic} shows a different arrangement than ISO/IEC 25010. Here, availability is part of security and not reliability. And so on and so forth.

To get a feeling about the effects of increasing dimensionality, just consider the relation between the two volumes of the n-dimensional unit-sphere $V_{unit\,sphere}$ and the n-dimensional cube $V_{cube}$ which comprises the unit-sphere. From an intuitive point of view, all points in the sphere are similar in the sense that their distance towards the center is smaller than 1.  

\begin{equation}
V_{unit\, sphere} = \frac{\pi^{n/2}}{n\Gamma(\frac{n}{2} +1)}, \, 
V_{cube} = 2^n
\end{equation}

\begin{figure}[ht]
\begin{center}
\includegraphics[width=8cm]{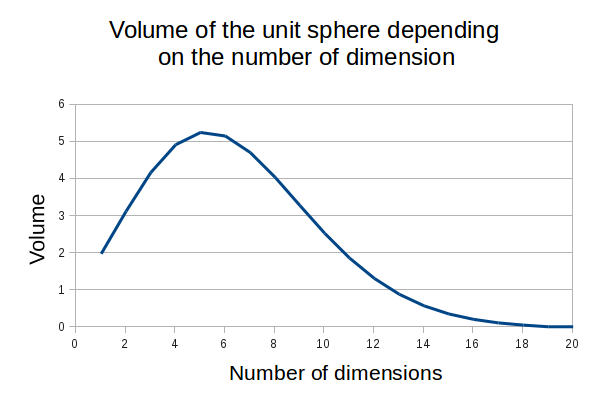}\\
\includegraphics[width=8cm]{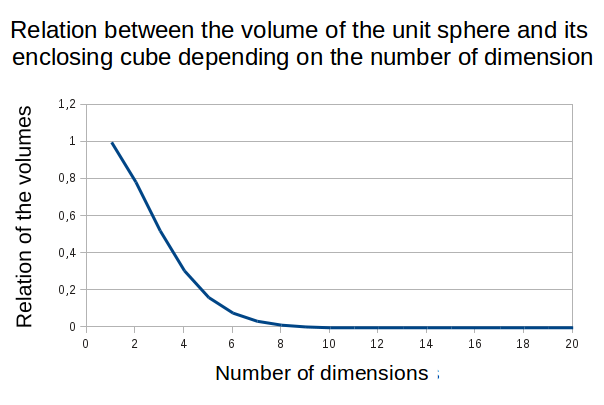}
\end{center}
\caption[]{\label{fig_volume_relation_sphere_to_cube} The behavior of the volume of the unit sphere and the relation between the volumes of unit sphere and unit cube for increasing dimensionality.}
\end{figure}

As is shown in Fig. \ref{fig_volume_relation_sphere_to_cube}, the relation of both volumes converges pretty quickly to zero.
\begin{equation}
\lim_{d\rightarrow \infty}\frac{V_{unit\,sphere}}{V_{cube}} = 0
\end{equation}
Intuitively one could say that with increasing dimensionality, most of the space appears ''in the edges'' far away from the center, or, most of the points will intuitively be dissimilar. While the length of the vectors to the sphere remain 1 with increasing dimensionality, the length of the vectors to the edges of the $n$-dimensional cube increase indefinitely.

\begin{figure}[ht]
\begin{center}
  \includegraphics[width=8cm]{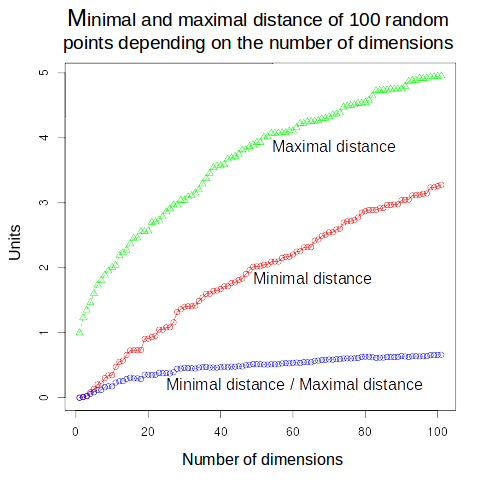}
\end{center}
\caption[]{\label{fig_numerical_simulations} A numerical simulation for 100 data points showing the minimal, maximal distance and their relation for adding increasing dimensions with random values.}
\end{figure}

Even more surprisingly, under certain quite robust assumptions on the distribution of data points, the distance function looses its usefulness with increasing dimensionality as the minimum and the maximum distance of data sets get closer and closer with increasing dimensionality (see Fig. \ref{fig_numerical_simulations}).

\begin{equation}
\lim_{n\rightarrow \infty}\frac{dist_{max}}{dist_{min}} = 1
\end{equation}

The core question is: how can we deal with this ''curse of dimensionality''? Under the headline of ''high dimensional data analysis'', there exists an enormous amount of literature on this topic and it would go far beyond the scope of this article to provide an overview on the diverse approaches.

Instead, I would like to offer a simple but nevertheless often effective heuristic approach, which is to assume, that for a system under consideration, there are few, very significant relations to its environment and many less significant ones. That is, we do not look solely at the values a quality dimension can take, but weight it with its significance. If the different quality dimensions can be sorted by their significance and this significance converges quickly enough to zero then we can sort the quality dimensions into two different categories.  

If we look at a customer who wants to by a product, we must relate the significance to the buy-decision of the customer. The first category are the ''strategic qualities''. There are usually only a handful of them. They generate contrast that differentiate a product positively and whose presence promote a buy-decision of a customer. One could interpret Rosser Reeves' ''Unique selling proposition'' \cite{Reeves1961} as relating to such key qualities. He characterized a unique selling proposition as a proposition which convinces the potential customer by pointing to a specific benefit and which the competition cannot or does not offer. 

The other category could be called ''necessary qualities''. These are the rest. They tend to reduce the contrast, because it is more their lacking that would attract attention, not so much their presence which is more or less taken for granted. They differentiate a product rather negatively and from an economic perspective, it is their absence that promotes a ''don‘t buy'' decision.

\subsection{Compromising on quality}
Generally, and in contrast to the unequivocal definition of functionality, the realization of different quality dimensions usually requires some sort of compromise – a system can not excel all requirements of all qualities simultaneously. For example, realizing low price and longevity at the same time is often not possible. Other examples are memory efficiency versus CPU efficiency, robustness versus extensibility, etc. The distinction between strategic qualities and necessary qualities can guide the choice of which quality to give greater credence to.

\subsection{Quality as a value}
As demonstrated, the relations of most systems to the rest of the world is much too complex to be comprehended completely by quantifiable, measurable parameters. Dealing with this kind of incomplete information is a very common situation for humans and we have developed a lot of different techniques to cope with which I would like to subsume under the notion of ''values''.

A value in this sense is a general action rule for a situation where only incomplete information is available to decide upon some actions and where a good-enough solution is sufficient. This lack of information may be due to principal reasons or just because it is uneconomical in the sense that it costs more to gain the information than could be gained with its acquisition. Another term for ''value'' in this sense could be ''action-heuristics''.

For example, one is generally well advised to behave polite or even kind towards our fellow humans. In every encounter, it is unclear, whether and under what circumstances we will meet again. Because of this incomplete information and the really low costs of behaving polite, it makes sense to behave polite under most circumstances without trying to evaluate each situation in particular.

Because of the general nature of values, it is to be expected that they will contradict each other at least partly. Being open and honest and polite all at the same time might not be possible. We then have to decide which value or action-heuristic is more appropriate. This could be done in a static hierarchy, which would be an action-heuristic by itself, a meta-value so to speak. Or it could also be a case-by-case decision.

Dealing with insufficient information is a huge subject on its own and an elaborate discussion goes far beyond the scope of this article. I just want to mention Herbert Simon who introduced the concept of bounded rationality \cite{Simon1982}, showing human decision making to develop by adapting to the environment to make well enough or satisficing and not optimal decisions - as the traditional concept of rationality assumed these days. Daniel Kahnemann and Amos Tversky \cite{KahnemanTversky1979} described by their ''prospect theory'' how humans make decisions under risk based on the estimated value of gains and losses using certain heuristics rather than the final outcome as was assumed by traditional decision theory. Gerd Gigerenzer and his Adaptive Behavior and Cognition (ABC) Group \cite{Gigerenzer2011_Heuristics} emphasizes that heuristics are indispensable and often even more(!) accurate than complex methods. 

The use of heuristics points to the enormous influence of learning and experience in taking satisficing decisions when compromising on quality. 
Is is evident that especially literature on quality for management usually underlines the value-aspect (e.g. \cite{Juran1951_Quality}) emphasizing that quality has to be supported by the top level leadership and has to be part of the company culture.

%
\section{Summary}
%

To sensibly approach quality in product development we have to be aware of what we know and what we don't know. The former means to weigh different possibilities and take some informed decisions while the latter requires a fall back on some general action rules. Where our knowledge ends, experience-based heuristics must fill the gap.

Yes, quality should be objectified by measurement. But there are two important, unavoidable hindrances: first, there is an inherent conflict between the narrowness of the considered context, which improves the validity of the measure but at the same time reduces its applicability to other contexts.
Second, any quality measurement is inherently incomplete. Due to the conflict between the many important quality dimensions and the necessary exemplary approach to quality, we have to find an appropriate antidote against the ''curse of dimensionality''. One simple but often effective heuristic is to distinguish between contrast generating strategic or key qualities and contrast diminishing necessary qualities. While a product should excel in the key dimensions, a ''good enough'' approach with the necessary dimensions suffice. This helps to structure discussions about business models.

A strategic approach towards quality for product development requires deep business as well as technical knowledge and experience. Hence, a tight cooperation between the business as well as the technical experts becomes necessary. Quality measurement helps in reaching a common understanding. Many usually tacit assumptions have to be made explicit. As measurements ideally are reproducible and unequivocal, the discussions of quality issues are shifted to a higher intellectual level towards questions of context and relevance. But still, it remains a delicate discussion. Understanding the principal limitations of measuring quality, companies should avoid mechanisms that exerts an unfavorable influence on this discussion. Otherwise they will realize - as VW had to with its recent diesel scandal - that their smart employees can use numbers all too well to disguise their individual interests as the desired goals of the company. \\[0.5cm]

\section{Acknowledgments}
My deep gratitude goes to Andreas Blumenthal who read a preliminary version of the manuscript and made a lot of very valuable suggestions, although we still debate the question how much simplification sticks in viewing functionality and quality as separable and being strictly complementary. I gratefully thank Peter Emmel who organized the innovation engineering lecture series at the DHBW Mannheim where most of the presented ideas have been developed for. I also kindly thank Jochen Boeder who encouraged me to cast these and other ideas into an e-learning session for SAP product experts and architects which I used as a starting point for this article and who provided me important feedback. I am also indebted to Mark Crawford who supervised my English of the e-learning.

\bibliographystyle{abbrv} 
\bibliography{soziologie,philosophy,informatics}

\begin{thebibliography}{10}

\bibitem{AlQutaish2010_quality}
R.~E. Al-Qutaish.
\newblock Quality models in software engineering literature: an analytical and
  comparative study.
\newblock {\em Journal of American Science}, 6(3):166--175, 2010.

\bibitem{AvizienisLaprieRandellLandwehr2004_Basic}
A.~Avizienis, J.-C. Laprie, B.~Randell, and C.~Landwehr.
\newblock Basic concepts and taxonomy of dependable and secure computing.
\newblock {\em Dependable and Secure Computing, IEEE Transactions on},
  1(1):11--33, 2004.

\bibitem{Bellman1961}
R.~E. Bellman.
\newblock {\em Adaptive Control Processes}.
\newblock Princeton University Press, NJ., 1961.

\bibitem{BoehmEtAl1978_Characteristics}
B.~W. Boehm, J.~R. Brown, H.~Kaspar, M.~Lipow, M.~G. J, and M.~M. J.
\newblock {\em {Characteristics of software quality}}.
\newblock TRW Softw. Technol. North-Holland, Amsterdam, 1978.

\bibitem{CandidoSantos2011_TQM}
C.~J. C{\^a}ndido and S.~P. Santos.
\newblock {Is TQM more difficult to implement than other transformational
  strategies?}
\newblock {\em Total Quality Management \& Business Excellence},
  22(11):1139--1164, 2011.

\bibitem{ChungLeite2009}
L.~Chung and J.~C.~S. do~Prado~Leite.
\newblock On non-functional requirements in software engineering.
\newblock In A.~Borgida, V.~K. Chaudhri, P.~Giorgini, and E.~S.~K. Yu, editors,
  {\em Conceptual Modeling: Foundations and Applications}, volume 5600 of {\em
  Lecture Notes in Computer Science}, pages 363--379. Springer, 2009.

\bibitem{GieslHafstein2015}
P.~Giesl and S.~Hafstein.
\newblock Review on computational methods for lyapunov functions.
\newblock {\em Discrete and Continuous Dynamical Systems - Series B (DCDS-B)},
  20(8):2291 -- 2331, 2015.

\bibitem{Gigerenzer2011_Heuristics}
G.~Gigerenzer, R.~Hertwig, and T.~Pachur.
\newblock {\em Heuristics: The foundations of adaptive behavior}.
\newblock Oxford University Press, Inc., 2011.

\bibitem{Gilb1976_Software}
T.~Gilb.
\newblock {\em Software Metrics}.
\newblock Winthrop Publishers, Inc., Cambridge, 1976.

\bibitem{IEC_60050}
{IEC 60050 International Electrotechnical Vocabulary}, 2001ff.

\bibitem{IEEE_1061-1998}
{IEEE 1061-1998 Standard for a Software Quality Metrics Methodology}, 1998.

\bibitem{ISO_IEC_25000}
{ISO/IEC 25000:2014 Systems and software engineering -- Systems and software
  Quality Requirements and Evaluation (SQuaRE) -- Guide to SQuaRE}, 2014.

\bibitem{ISO_IEC_25010}
{ISO/IEC 25010:2011 Systems and software engineering -- Systems and software
  Quality Requirements and Evaluation (SQuaRE) -- System and software quality
  models}, 2011.

\bibitem{ISO_IEC_25020}
{ISO/IEC 25020:2007 Software engineering -- Software product Quality
  Requirements and Evaluation (SQuaRE) -- Measurement reference model and
  guide}, 2007.

\bibitem{ISO_IEC_9126}
{ISO/IEC 9126:2001 Software engineering -- Product quality -- Part 1: Quality
  model}, 2001.

\bibitem{Juran1951_Quality}
J.~M. Juran.
\newblock {\em Quality Control Handbook}.
\newblock McGraw-Hill, New York, 1951.
\newblock 6th edition, 2010.

\bibitem{KahnemanTversky1979}
D.~Kahneman and A.~Tversky.
\newblock Prospect theory: An analysis of decision under risk.
\newblock {\em Econometrica}, 47:263--291, 1979.

\bibitem{KanerBond2004_Software}
C.~Kaner and W.~P. Bond.
\newblock Software engineering metrics: What do they measure and how do we
  know?
\newblock In {\em 10th International Software Metrics Symposium, Metrics 2004},
  pages 1--12, 2004.

\bibitem{KatzLindell2015}
J.~Katz and Y.~Lindell.
\newblock {\em Introduction to Modern Cryprography}.
\newblock Chapman \& Hall/CRC, 2 edition, 2015.

\bibitem{McCallRichardsWalters1977}
J.~A. McCall, P.~Richards, and G.~Walters.
\newblock Factors in software quality, vol 1, 2, 3.
\newblock Technical Report RADC-TR-77-369, General Electric Company, 1977.

\bibitem{Mosadeghrad2012_Towards}
A.~M. Mosadeghrad.
\newblock Towards a theory of quality management: an integration of strategic
  management, quality management and project management.
\newblock {\em International Journal of Modelling in Operations Management},
  2(1):89--118, 2012.

\bibitem{Reeves1961}
R.~Reeves.
\newblock {\em Reality in Advertising}.
\newblock Knopf, New York, 1961.

\bibitem{Reich2001}
J.~Reich.
\newblock Über {S}truktur oder das {V}erhältnis der {T}eile zum {G}anzen.
\newblock {\em Philosophia Naturalis}, 38(1):37--69, 2001.

\bibitem{Reich2016_systems}
J.~Reich.
\newblock Composition, cooperation, and coordination of computational systems.
\newblock {\em CoRR}, abs/1602.07065, 2016.

\bibitem{Sifakis2011_Vision}
J.~Sifakis.
\newblock A vision for computer science - the system perspective.
\newblock {\em Central European Journal of Computer Science}, 1(1):1008--116,
  2011.

\bibitem{Simon1982}
A.~Simon, Herbert.
\newblock {\em Models of bounded rationality}.
\newblock MIT Press, Cambridge, MA, 1982.

\bibitem{SousaVoss2002_Quality}
R.~Sousa and C.~A. Voss.
\newblock Quality management re-visited: a reflective review and agenda for
  future research.
\newblock {\em Journal of operations management}, 20(1):91--109, 2002.

\bibitem{Bertalanffy1968_GST}
L.~von Bertalanffy.
\newblock {\em {General System Theory: Foundations, Development,
  Applications}}.
\newblock George Braziller, New York, 1968.

\bibitem{WagnerEtAl2012_Software}
S.~Wagner, K.~Lochmann, S.~Winter, A.~Goeb, M.~Kläs, and S.~Nunnenmacher.
\newblock Software quality models in practice.
\newblock Technical Report TUM-I129, Technische Universität München, 2012.

\end{thebibliography}

\end{document}